\documentclass[prl,twocolumn,showpacs,preprintnumbers,amsmath,amssymb]{revtex4}
\usepackage{dcolumn}
\usepackage{bm}
\usepackage{graphicx}
\usepackage{bm}
\begin{document}
\title{Critical exponents and the correlation length in the charge exchange manganite spin glass  Eu${_{0.5}}$Ba${_{0.5}}$MnO${_3}$}
\author{Sunil Nair}
\email{sunilnair@tifr.res.in}
\author{A. K. Nigam}
\affiliation{Department of Condensed Matter Physics and Material Science,\\Tata Institute of Fundamental Research,\\ Homi Bhabha Road, Mumbai 400 005, INDIA.}
\date{\today}
\begin{abstract}
The critical regime of the charge exchange (CE) manganite spin glass  Eu${_{0.5}}$Ba${_{0.5}}$MnO${_3}$ is investigated using linear and non linear magnetic susceptibility and the divergence of the third ordered susceptibility ($\chi{_3}$) signifying the onset of a conventional freezing transition is experimentally demonstrated. The divergence in $\chi{_3}$, dynamical scaling of the linear susceptibility and relevant scaling equations are used to determine the critical exponents associated with this freezing transition, the values of which match well with the 3D Ising universality class. Magnetic field dependence of the spin glass response function is  used to estimate the spin correlation length which is seen to be larger than the charge/orbital correlation length reported in this system.
\end{abstract}
\maketitle 
The hole doped mixed valent manganites of the type AMnO${_3}$ have been extensively investigated on account of its diverse phase diagram which in turn arises due to a complex interplay between the lattice, spin, charge and orbital degrees of freedom. Though earlier emphasis has been on the study of phenomenon like collosal magnetoresistance and charge/orbital ordering (CO-OO)\cite{rao}, more recent investigations have brought forward the phenomenon of \emph{glassiness} in a variety of systems\cite{ps}. Observation of these phenomenon, which are normally associated with a conventional spin glass like freezing phase transition has brought to the forefront the issue of whether \emph{glassy} manganites are truly glasses (and thus can be dealt with using conventional spin glass theories) or whether they constitute a fundamentally different state. This issue is further complicated due to the effects of electronic phase separation (EPS) which result in the system subdividing itself into self organised regions with varying hole concentrations\cite{dag1}. This EPS could either result in formation of regions with competing magnetic interactions or in self generated clusters which can exhibit superparamagnetic like \emph{blocking} at a well defined temperature\cite{sunil}. Thus, the fundamental problem here pertains to unambigously determining whether the observed glassy behavior is purely dynamic (where the average relaxation time of the system exceeds the timescales of the experimental probes) or on the contrary, it is associated with an underlying thermodynamic transition as in conventional spin glasses.

The prerequisite for a cooperative glassy state is the presence of disorder (either site or bond) which prevents the stabilisation of a long range ordered magnetic ground state\cite{mydosh}. The same is true within the phase separation scenario, where disorder promotes the stabilisation of self generated clusters, interaction between which result in glass like experimental signatures\cite{riva}. Not surprisingly, most manganite systems which exhibit glassy behaviour have inherent quenched disorder, either by the direct destabilisation of the Mn-O-Mn network through Mn site substitution\cite{veglio} or the random potential which arises from the electrostatic or lattice disorder due to ionic size mismatch between the A site ions\cite{maig}. The most drastic effect of quenched disorder is seen in narrow bandwidth systems near half doping\cite{ce}, which stabilise in the \emph{charge exchange}(CE) type of antiferromagnetic ordering which is made up of a checkerboard arrangement of Mn${^{3+}}$ and Mn${^{4+}}$ ions, accompanied by an associated orbital ordering at the Mn${^{3+}}$ sites\cite{goodenough}. The non trivial geometrical arrangement of the zigzag chains makes it extremely susceptible to disorder, and it is well known that Mn site disorder results in a destabilisation of the long range magnetic order leading to magnetic ground states as diverse as relaxor ferromagnets\cite{kimura} or superparamagnetism\cite{sunil}. Recently, it has been reported that in narrow bandwidth systems, quenched disorder in the form of large lattice mismatch between the A site ions can result in a complete supression of the CO-OO state resulting in a \emph{charge exchange spin glass}\cite{tok1}. Interestingly, this state is not associated with macroscopic electronic phase separation, which implies that the spin glass is nearly atomistic\cite{tok2}, thus making it very attractive as far as investigation of clean manganite glasses are concerned. 

The absolute characterisation of any phase transition involves the accurate determination of the critical exponents (related to the power law singularities in various physical parameters) which enable the classification of the phase transition into well established universality classes\cite{stanley}. In the context of spin glasses, the appropriate parameter to measure is the nonlinear susceptibility which by virtue of being an indicator of the Edwards Anderson order parameter, diverges at the freezing temperature\cite{binder}. Direct measurement of the nonlinear susceptibility has the added advantage of enabling one to unambigously distinguish between a spin glass like freezing from a superparamagnetic like blocking, since in superparamagnets, the third order susceptibility ($\chi{_3}$) does not diverge and is also known to exhibit a T${^{-3}}$ dependence above the blocking temperature\cite{ash}. In this letter, we report the first investigation of the critical regime of a perovskite manganite glass with an aim to determine the critical exponents and thus the universality class associated with this transition. Since the CE glass is thought to arise from a homogenous short range charge/orbital order, the spin glass correlation length is estimated in order to verify the validity of the classical Goodenough picture of identical magnetic and orbital correlation lengths. All the measurements are done on a ceramic sample of Eu${_{0.5}}$Ba${_{0.5}}$MnO${_3}$  prepared by the solid state route. Bulk magnetic measurements are done using a home made ac susceptometer and commercial (Quantum Design SQUID and Oxford Maglab) magnetometers. 

Fig. 1 shows the magnetic history effect as measured in Eu${_{0.5}}$Ba${_{0.5}}$MnO${_3}$ clearly indicating a pronounced irreversibility at T${_g}$ $\approx$ 42K. The inset shows the MH isotherm as measured at 5 K indicating a finite loop width with no trace of saturation upto the highest measured field. The lower panel shows the frequency dependence of the imaginary part of ac susceptibility ($\chi{^{''}}$) which indicates a enhancement in the peak magnitude as well as a shift in T${_g}$ as the frequency of measurement is reduced. Assuming conventional critical slowing down on approaching T${_g}$ from the high-T side, the relaxation time $\tau$ can be expressed as $\tau = \tau^{*}(T/T_{g} -1 )^{-z\nu}$; where $\tau$ is (2$\pi$f)$^{-1}$ with f being the measurement frequency, $\tau^{*}$ is the microscpic flipping time of the fluctuating entities, z is the dynamical critical exponent and $\nu$ is the critical exponent associated with the spin correlation length $\xi$. This is shown in the inset with the solid line being the fit giving a value of $z\nu$ = 7.1 $\pm$ 0.2 which matches well with that reported earlier on single crystal specimens\cite{tok2}. 

By virtue of being proportional to the four spin correlation function, the third ordered susceptibility $\chi_{3}$ exhibits a power-law crtical divergence at the spin glass freezing temperature T${_f}$ of the form $\chi_{3}$ $\propto$ $\epsilon^{-\gamma}$, where the reduced temperature $\epsilon$=(T-T${_f}$)/T${_f}$ and $\gamma$ is the critical exponent characteristic of a phase transition to a spin glass state. Thus, though the H and T dependendent peak in the linear susceptibility ($\chi_{1}$) is non divergent, the higher order susceptibility $\chi{_3}$ diverges in the limits H$\rightarrow$0 and T$\rightarrow$T${_g}$. This is shown in Fig. 2, using the third ordered susceptibility ($\chi_{3}$)  data measured using ac susceptibility \cite{ft0}. The inset shows the log-log plot of $\chi_{3}$ as a function of the applied field H showing a true critical divergence. The main panel shows a log-log plot of $\chi_{3}$ as a function of the reduced temperature $\epsilon$. The straight fit gives a value of the critical exponent $\gamma$ = 2.74$\pm$ 0.05. To the best of our knowledge, this is the first report of the critical divergence of the non linear susceptibility with H and T in any manganite spin glass, thus confirming that the low temperature ground state in the system Eu$_{0.5}$Ba$_{0.5}$MnO$_3$ arises from a freezing of the spin degrees of freedom.

Dynamical scaling of the ac susceptibility $\chi^{'}(\omega, T) + i\chi^{''}(\omega, T)$ has been extensively used as supporting evidence for critical behavior associated with phase transitions in frozen magnetic systems. A common route is to use the scaling equation $\chi^{''}T/\omega^{\beta/{z\nu}} \approx g(t/\omega^{1/{z\nu}})$, where g is the scaling function\cite{huse}. A linear scaling plot of this form, where $\chi^{''}T/\omega^{\beta/{z\nu}}$ is plotted against $t/\omega^{1/{z\nu}}$ is shown in Fig. 3, using the $\chi_{1}^{''} (\omega,T)$ data as obtained in Eu${_{0.5}}$Ba${_{0.5}}$MnO${_3}$. To minimise the probablity of erroneous scaling arising from the presence of a large number of free running variables, only the value of $\beta$ was varied during the course of this scaling procedure. The best scaling was obtained for a value of $\beta$ = 0.6 $\pm$ 0.1, with all the peaks in $\chi^{''}T$ coalescing on the same point. 

With the value of the exponents z$\nu$, $\gamma$ and $\beta$ thus obtained, the other exponents can be estimated using the scaling equations $\alpha$ + 2$\beta$ + $\gamma$ = 2,d$\nu$ = 2-$\alpha$ and $\delta$ = $\gamma$/$\beta$ + 1, where d refers to the dimensionality and is taken to be 3 in this case. The values of the exponents thus determined is tabulated in table 1. The values of the exponents as expected for a 3D Ising system from MC simulations\cite{x} and that experimentally determined for a well established 3D Ising system Fe${_{0.5}}$Mn${_{0.5}}$TiO${_3}$ \cite{gunnar}is also shown for the sake of comparison. As is clearly seen, the values of the exponents determined for the system Eu${_{0.5}}$Ba${_{0.5}}$MnO${_3}$ matches very well with that expected theoretically and in some cases represents an improvement over prior experimental reports, thus unambigously indicating that the spin glass transition in this manganite system belongs to the 3D Ising universality class. In this context, it is interesting to note that the exponents measured on the ferromagnetic compositions of various hole doped manganites have indicated that the double exchange driven transitions belong to the isotropic Heisenberg universality class\cite{sunil1}. The effective anisotropy in our case which results in a Ising like exponents most probably arises as a consequence of short range orbital ordering which is endemic to narrow bandwidth half doped systems in the presence of disorder. Recently, a layered manganite spin glass was reported to exhibit exponents which lie between the Heisenberg and Ising universality classes \cite{mat}, presumably because the short range orbital order within the layered structure mainly includes 3x${^2}$-r${^2}$/3y${^2}$-r${^2}$ orbitals favoring inplane magnetic moments.
\begin{table}
\caption{\label{tab:table 1}Critical exponents of the perovskite manganite Eu${_{0.5}}$Ba${_{0.5}}$MnO${_3}$ as determined using linear and nonlinear susceptibility data. The values of the exponents as expected theoretically in a 3D Ising system \cite{x} and that reported in a well established Ising system Fe${_{0.5}}$Mn${_{0.5}}$TiO${_3}$ \cite{gunnar} is also given for the sake of comparison.}

\begin{tabular}{c|c|c|c}
\hline
Exponent &3D Ising  &Eu${_{0.5}}$Ba${_{0.5}}$MnO${_3}$ &Fe${_{0.5}}$Mn${_{0.5}}$TiO${_3}$\\
\hline
$\gamma$ &2.9$\pm$0.3 &2.74$\pm$0.05 &4.0$\pm$0.3 \\
$\beta$ &0.5 &0.6$\pm$0.1 &0.54 \\
z &6.0$\pm$0.8 &5.4$\pm$0.5 &6.2 \\
$\nu$ &1.3$\pm$0.1 &1.31$\pm$0.08 &1.7 \\
$\alpha$  &-1.9$\pm$0.3 &-1.9$\pm$0.25 &-3.1 \\
$\delta$ &6.8$\pm$0.6 &5.7$\pm$0.9 &8.4$\pm$1.5 \\
\hline
\end{tabular}
\end{table}

The spin glass relaxation rate S (defined as S(t) = d[-M${_{trm}}$(t,t${_w}$)/H]/d\emph{ln}t; where M${_{trm}}$(t,t${_w}$) is the thermoremanent magnetisation at time \emph{t} after cutting the magnetic field to zero) is directly related to the typical value of the free energy barriers which can be explored on the available experimental time scales\cite{vincent}. For any given waiting time (t${_w}$), local equilibration of the state occupancies result in a peak in this response function for measurement times in the vicinity of t${_w}$. The magnetic field dependence of this peak has been used to estimate the volume over which the spins are effectively locked together for barrier hopping, and thus can be used to estimate the spin glass correlation length\cite{joh}. The M${_{trm}}$ as measured in the sample Eu${_{0.5}}$Ba${_{0.5}}$MnO${_3}$ is shown in Fig. 4. All measurements were done at 0.7T${_G}$ after cooling the sample in the presence of a field H from 2.5T${_G}$. Prior to cuting off the field, the system was made to wait for a time t${_w}$$\approx$60 secs. The inset shows the magnetic field dependence of S clearly indicating that the measuring time at which S(t) peaks reduces as a function of the magnetic field H. Equating this change to be associated with a magnetic energy E${_z}$, for small values of which the apparent age t${_{eff}}$ (smaller than the actual waiting time t${_w}$) is given by lnt${_{eff}}$/t${_w}$ = -E$_{z}$/k${_B}$T. As was done before for Ising systems\cite{bert}, this Zeeman energy term E${_z}$ can be equated to  $\sqrt{N}H(m\mu_{B})$, where N is the number of spins which are effectively blocked together and $m\mu{_B}$ is the effective moment of one spin entity. Interestingly, the value of N thus calculated in this system is seen to be of the order of 10${^5}$\cite{ft1}, which is larger than that reported for the Ising system Fe${_{0.5}}$Mn${_{0.5}}$TiO${_3}$ and comparable with that reported in Heisenberg glasses\cite{bert}. The effective correlation length ($\xi_{mag}$) can then be estimated to be of the order of 35 $\AA$\cite{ft2}, which is larger than the charge-orbital correlation length ($\xi_{orb}$) estimated for this system\cite{tok2} by a factor of 1.75. Interestingly, Soft X-Ray resonant experiments on a COO manganite manganite near half doping have shown that the magnetic correlation length exceeds that of the orbital order by a factor of 2\cite{thomas}. This challenges the classical Goodenough model where orbital domain walls create magnetic domain walls in the Mn${^{3+}}$ sublattice which in turn would imply that the ratio of $\xi_{mag}$/$\xi_{orb}$ is unity. Our estimates of $\xi_{mag}$ would indicate that a ratio of $\xi_{mag}$/$\xi_{orb}$ $\approx$ 2 is valid not only for OO systems with concommitant long range magnetic order, but also in systems where disorder results in a suppression of magnetic long range ordering. The temperature (T$^{*}$)\cite{burgy}, where these short range correlations in this system would form is $\approx$ 2.5T${_G}$ as can be estimated from the deviation from linearity of $\chi^{-1}$ in the paramagnetic regime.
 
In summary, we have investigated the critical regime of the charge exchange spin glass Eu${_{0.5}}$Ba${_{0.5}}$MnO${_3}$ using linear and nonlinear ac susceptibility. The critical divergence of the third ordered susceptibility ($\chi{_{3}}$) as a function of H and T is demonstrated, thus confirming that the low temperature metastable magnetic phase arises from the freezing of the spin degrees of freedom. The exponents determined unambigously show that this transition falls into the 3D Ising universality class. The spin glass correlation length is estimated from magnetic field change aging experiments and indicates that this magnetic correlation length is larger than the charge/orbital length reported earlier. Keeping in mind the fact that recent experiments have challenged the basic premises of the classical Goodenough model of CE systems, our results calls for more systematic investigations of the charge/orbital and magnetic length scales in such systems.

\begin{figure}
\caption{The upper panel shows the bulk DC magnetisation as measured in Eu${_{0.5}}$Ba${_{0.5}}$MnO${_3}$ in the field cooled and zero field cooled cycles, with the inset depicting the MH isotherm as measured at 5K. The lower panel exhibits the frequency dependence of the imaginary part of the linear susceptibility for the same sample. The inset shows the dynamical scaling of $\tau$(T${_f}$) with the reduced temperature $\epsilon$ giving T${_G}$ = 41.04$\pm$0.07 and z$\nu$ = 7.1$\pm$0.2.}
\caption{log-log plot of $\chi{_3}$ as a function of the reduced temperature ($\epsilon$) indicating the critical divergence of $\chi_3$ in the limit T$\rightarrow$T${_G}$. The straight line fit gives the susceptibility exponent $\gamma$ = 2.74 $\pm$ 0.05. The inset shows the log-log plot of $\chi{_3}$ as a function of the applied field H  indicating the divergence of $\chi{_3}$ in the limit H$\rightarrow$0.}
\caption{A linear scaling plot of $\chi^{''}T/\omega^{\beta/{z\nu}}$ plotted against $t/\omega^{1/{z\nu}}$ using $\chi^{''}_{1}$(T) data giving the value of $\beta$ = 0.6 $\pm$ 0.1}
\caption{M${_{TRM}}$ as measured at 0.7T${_G}$ after cooling in the presence of a field of 100 Oe and waiting (t${_w}$) for 60 secs. The inset shows the magnetic field dependence of the relaxation rate S(t) at applied fields of 100, 150 and 200 Oe.}
\end{figure}

\begin{thebibliography}{ }
\bibitem[26]{ft1}Here N (and thus $\xi{_{mag}}$) is a function of both the waiting time (t${_w}$) and the applied field (H); the exact functional forms of which are presently being investigated. For the sake of comparison with other lengthscales of importance like $\xi{_{orb}}$, the value quoted is that obtained for the lowest H (100 Oe) and t${_w}$(60 secs) from among our datasets.
\bibitem[27]{ft2}This is estimated by considering a pseudocubic lattice of 3.8$\AA$ as was done earlier\cite{tok2} and by assuming spherically symettric clusters of N unit cells.
\end{thebibliography}
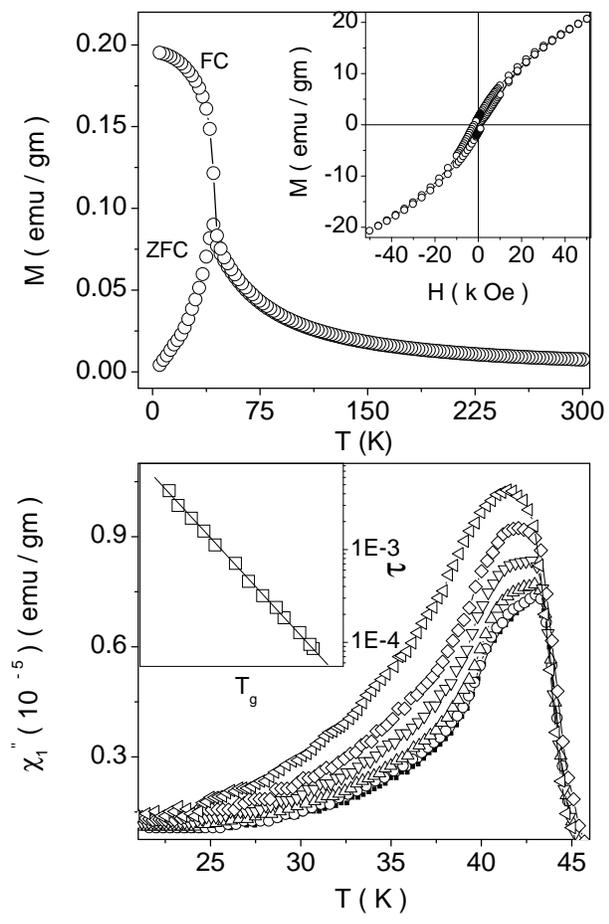
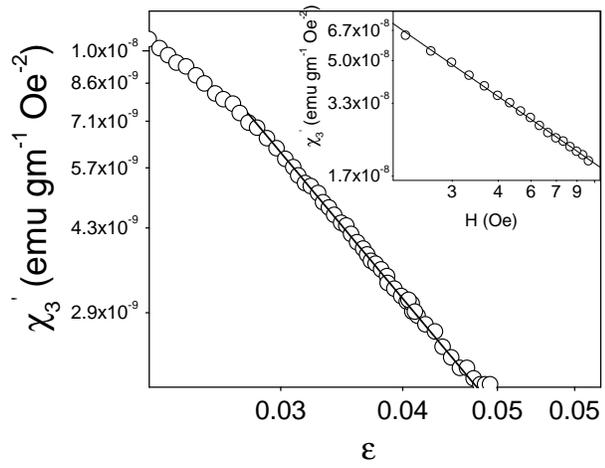
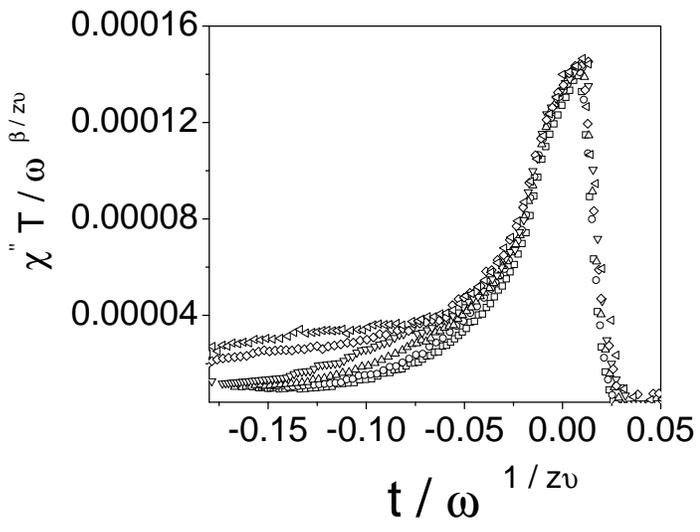
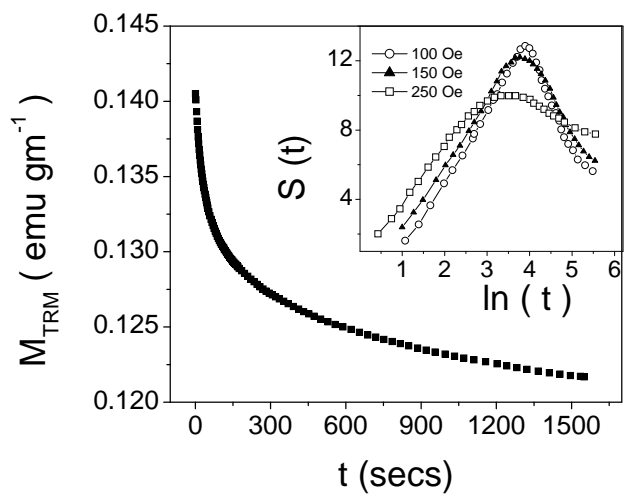
\end{document}